\documentclass[conference,fleqn]{IEEEtran}
\usepackage{amssymb}
\usepackage[tight,TABTOPCAP]{subfigure}
\usepackage{amsfonts}
\usepackage[dvips]{graphicx}
\ifCLASSINFOpdf
\else
\fi
\usepackage[cmex10]{amsmath}
\begin{document}
%
\title{Extended Range Profiling in Stepped-Frequency Radar with Sparse Recovery}

\author{\IEEEauthorblockN{\emph{Yang Hu, ~Yimin Liu, ~Huadong Meng, ~Xiqin Wang}}
\IEEEauthorblockA{Department of Electronic Engineering,
~Tsinghua University,~Beijing\\
Email: foxsheep@gmail.com}

}

\maketitle

\begin{abstract}
The newly emerging theory of compressed sensing (CS) enables
restoring a sparse signal from inadequate number of linear
projections. Based on compressed sensing theory, a new algorithm of
high-resolution range profiling for stepped-frequency (SF) radar
suffering from missing pulses is proposed. The new algorithm
recovers target range profile over multiple coarse-range-bins,
providing a wide range profiling capability. MATLAB simulation
results are presented to verify the proposed method. Furthermore, we
use collected data from real SF radar to generate extended target
high-resolution range (HRR) profile. Results are compared with
`stretch' based least square method to prove its applicability.
\end{abstract}

\section{Introduction}

The range resolution of a radar system is determined by the
bandwidth of the transmitted signal. Stepped-frequency (SF) pulse
train obtains large signal bandwidth by linearly shifting,
step-by-step, the center frequencies of a train of pulses. It is
widely used in high-resolution radar systems and well documented in
the literature \cite{einstein},\cite{wehner}. In SF radars, the
`stretch' processing method \cite{wehner}, based on inverse discrete
Fourier transform (IDFT) technique, can acquire high-range
resolution (HRR) profiles with narrow instantaneous bandwidth and
low system complexity. However, SF radar suffers greatly from
missing-pulse problems due to interference or jamming impinging on
the receiver, since SF technique occupies a large bandwidth. While
some pulses missing are and hence must be discarded, the IDFT based
stretch processing will inevitably leads to high sidelobes, thus
undermining the profiling quality. Various methods have been
proposed to interpolate the missing data (see \cite{zhang} and
reference therein). Theoretical analysis and experience indicate
that the longer the signal interpolation length is, the larger the
interpolation error is. If the missing pulse number becomes bigger,
the performance of existed method will reduce rapidly \cite{zhang}.

Besides the missing-pulse problem, SF radar suffers from `ghost
image' phenomenon. This problem, mainly caused by range ambiguity
among adjacent `coarse-range-bins', is delicately addressed in
\cite{liu}, where the author solved the problem by least square (LS)
technique. But this method is applicable on the assumption that full
pulses are well received, and the foundation of it is still IDFT.
Therefore, missing-pulses also deteriorate the profiling results.
Recently, the new emerging theory of compressed sensing (CS)
\cite{candes},\cite{donoho} that achieves high resolution has been
widely used in radar applications \cite{baraniuk}. The main
advantage of this theory is that, with sub-Nyquist samples, sparse
signal can still be reconstructed perfectly. CS theory was
introduced in the signal processing for SF radar by Sagar Shah
\emph{et al.} \cite{shah}. With reduced number of transmitted pulses
in one coherent processing interval (CPI), their method provides
super-resolution ability in both range and Doppler domain. It also
indicted that missing-pulse problem can be solved with their method.
However, they only discussed profiling range of only one
coarse-range-bin, limiting their application on narrow-range-gate
profiling.

This paper introduces a new profiling algorithm for SF radar with
missing pulses, and the profiling range gate extents for multiple
coarse-range-bins. We focus on profiling of a stationary object.
Unavailable data from missing pulses are discarded; sparse recovery
is used to obtain extended synthetic range profile. We demonstrate
that new algorithm can solve the missing-pulse problem, it also has
a wide profiling range gate. The remainder of this paper is
organized as follows. In Section II, the signal model of HRR
profiling for SF radar is stated. CS based profiling with missing
pulses is described in section III. Simulation results are presented
in section IV. Section V concludes the new approach.

\section{System Model}
In SF radar, a pulse train of $N$ pulses are transmitted with
stepped carrier frequencies. For the $n$th pulse, the carrier
frequency is $f_n=f_c+n\Delta f$, where $f_c$ is the initial
frequency and $\Delta f$ the frequency step.
The complex profile of the measured scene can be represented by
system function $H(t_d)$, as has been derived in \cite{einstein}.
$t_d$ is the time domain variable, and $H(t_d)$ describes the
complex reflectivity of measured scene corresponding to time delay
$t_d$. For the convenience of signal modeling and derivation, it is
assumed that one target falls in the range gate [$R_0,R_0+D$] over
the whole coherent processing interval, where $R_0=cQ/2\Delta f$,
and $D=cL/2\Delta f$ ($Q$,$L$ are nonnegative integers and $c$ is
the speed of light). In the `stretch' processing \cite{einstein},
the range resolution is $c/2N\Delta f$ \cite{wehner}. Choosing this
resolution as the sampling period, the $p$th high-resolution range
cell, which represents the complex amplitude of the scatterer
located in the range $R_0+(cp/2N\Delta f)$, is written by
$h_p=H(p/N\Delta f)$. Thus, the HRR profile of the target can be
expressed by the vector $\textbf{h}=[h_0,h_1,…,h_{NL-1}]^T$.

A target response matrix (TRM) \cite{wehner} was used to organize
the echo signal of the pulse train. The TRM contains $N$ rows and
$S$ columns. The $n$th row consists of $S$ uniformly sampled
time-domain data from the baseband echo signal of the $n$th pulse
(If the transmitted baseband waveform is pulse compressing waveform,
the `baseband echo signal' refers to the pulse-compressed echo
signal). The elements in the same column are baseband samples of the
same coarse range cell. The column number is $S=2D/c\Delta t$, where
$\Delta t$ is the sampling interval. The TRM of a target can be
denoted by { \setlength\arraycolsep{0.5pt}
\begin{equation}
\textbf{E}=
\left[                 
\begin{array}{cccc}   
  E_0(0) & E_0(\Delta t) & \cdots & E_0(S\Delta t-\Delta t) \\  
  E_1(0) & E_1(\Delta t) & \cdots & E_1(S\Delta t-\Delta t) \\  
  \vdots & \vdots & \ddots & \vdots \\
  E_{N-1}(0) & E_{N-1}(\Delta t) & \cdots & E_{N-1}(S\Delta t-\Delta t) \\
\end{array}
\right].               
\end{equation}}
Here, $E_n(\tau)$ is the baseband echo signal of the $n$th pulse,
and $\tau$ is the baseband sampling instant. As derived in
\cite{einstein}, the baseband echo signal from stationary target is
\begin{equation}
E_n(\tau)=\sum_{p=0}^{NL-1}{h_p
R_X(\tau-\frac{2R_0}{c}-\frac{p}{N\Delta
f})e^{-j2\pi\frac{np}{N}}}+u_n(\tau)
\end{equation}
where $R_X(\tau)$ is the baseband pulse shape and $u_n(\tau)$ is
additive noise.

In the `stretch' processing method, the IDFT is applied to each TRM
column to form a HRR profile in one coarse-range-bin
\cite{einstein}. Missing pulse problem means data from some rows of
TRM are not available. If the missing pulse number is large,
profiling quality is greatly decreased using IDFT. Our new method
solve this problem by sparse recovery based on CS theory, which can
provide a better profiling quality. Based on the observation that
discrete system function vector $\textbf{h}$ is sparse, we propose a
new scheme for HRR profiling based on sparse recovery in the next
section.



\section{HRR profiling with missing pulses}
We now introduce the new CS based HRR profiling method, on condition
that some pulses are missing. Suppose only $M$ pulses ($M<N$) from
$N$ transmitted carrier frequencies are valid, that the carrier
frequency of the $m$th valid pulse is $F_m=f_c+C_m\Delta f$, where
$m$ is an integer between $0$ and $M-1$, $C_m$ is an integer between
$0$ and $N-1$. Substituting pulse number index $n$ in equation (2)
by $C_m$, we derive sample output for the $m$th pulse at sampling
instance $\tau$
\begin{equation}
E_{C_m}(\tau)=\sum_{p=0}^{NL-1}{h_p
R_X(\tau-\frac{2R_0}{c}-\frac{p}{N\Delta f})e^{-j2\pi\frac{C_m
p}{N}}}+u_m(\tau).
\end{equation}
We rewrite (3) in vector multiplication form:
\begin{equation}
E_{C_m}(\tau)=\varphi(C_m,\tau)\textbf{h}+u_m(\tau).
\end{equation}
$\varphi(C_m,\tau)$is a row vector of length $N\times L$, the $p$th
element of the vector is
\begin{equation}
\varphi_p(C_m,\tau)=R_X(\tau-\frac{2R_0}{c}-\frac{p}{N\Delta
f})e^{-j2\pi\frac{C_m p}{N}}.
\end{equation}

Deleting the invalid data in the TRM, the row number decreases to
$M$. {\setlength\arraycolsep{0.5pt}
\begin{equation}
\tilde{\textbf{E}}=
\left[                 
\begin{array}{cccc}   
  E_{C_0}(0) & E_{C_0}(\Delta t) & \cdots & E_{C_0}(S\Delta t-\Delta t) \\  
  E_{C_1}(0) & E_{C_1}(\Delta t) & \cdots & E_{C_1}(S\Delta t-\Delta t) \\  
  \vdots & \vdots & \ddots & \vdots \\
  E_{C_{M-1}}(0) & E_{C_{M-1}}(\Delta t) & \cdots & E_{C_{M-1}}(S\Delta t-\Delta t) \\
\end{array}
\right].               
\end{equation}}
The new TRM includes all available information we received. Note
that each element of the TRM is a linear projection of system
function $\textbf{h}$. By vectorizing this matrix, we may write the
following equation
\begin{equation}
Y=\text{vec}(\tilde{\textbf{E}})=\Phi \textbf{h}+U.
\end{equation}
The observation vector $Y$ is of length $M\times S$. Matrix $\Phi$
is the projection matrix of $M\times S$ rows and $N\times L$
columns, each row of $\Phi$ is corresponding to an observation. For
an instance, the row corresponding to pulse number $C_m$ and
sampling instance $s\Delta t$ is $\varphi(C_m,s\Delta t)$. $U$ the
noise vector for all observations. We have established a linear
projection for complex profile $\textbf{h}$. While pulses are
missing, $M<N$ holds, and inequality $MN<SL$ holds. Therefore, (7)
becomes an underdetermined equation. According to CS theory,
recovering a sparse signal from insufficient observation is possible
by $\ell_1$ minimization \cite{donoho}:
\begin{equation}
\min\|\tilde{\textbf{h}}\|_1  \;\;\;\;\;\; s.t.
\;\;\|Y-\Phi\tilde{\textbf{h}}\|_2\leq \epsilon
\end{equation}
where $\tilde{\textbf{h}}$ is an reconstruction of $\textbf{h}$ and
$\epsilon$ is an estimation error that is determined by received
signal noise.

\section{Results}
We show some primary results of simulation. The HRR range profile of
a real aircraft (Fig.1(a)) was measured by a wideband C-band chirp
radar. The chirp bandwidth was 512MHz, providing a range resolution
of about 0.3m. This measured range profile is used as the scatterer
truth. For SF radar simulation, 32 LFM pulses are transmitted in a
coherent pulse train. The frequency step size is 16MHz; and the
total effective bandwidth is $N\Delta f=$512MHz. In each pulse,
single-pulse bandwidth is 24MHz. The sampling rate $f_s$ equals
single-pulse bandwidth. The profile range gate covers $12$
coarse-range-bins. We simulate the missing pulse condition by
discarding data received from randomly selected 12 pulses, the left
20 pulses are valid. White Gaussian noise was added to the received
data, SNR is approximately 15dB.

\subsection{Simulated Data}
{\setlength{\abovecaptionskip}{-2pt}
\begin{figure}[]
\centering%
\subfigure []{
\includegraphics[width=0.4 \textwidth] {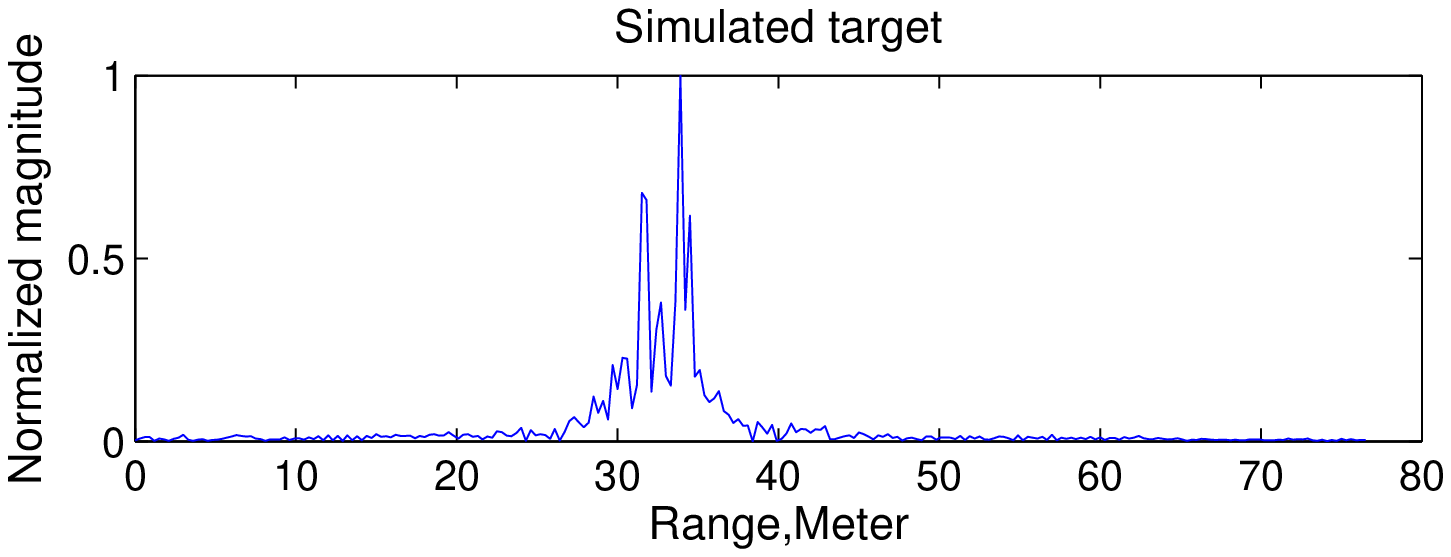}
} \subfigure []{
\includegraphics[width=0.4 \textwidth] {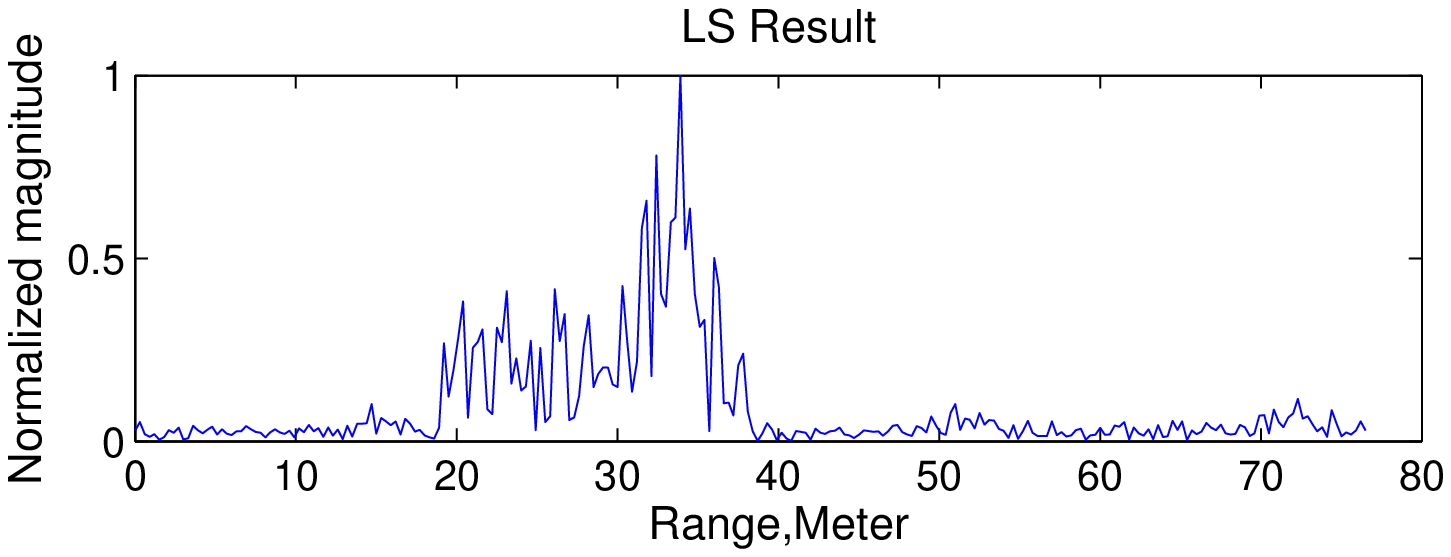}
} \subfigure []{
\includegraphics[width=0.4 \textwidth] {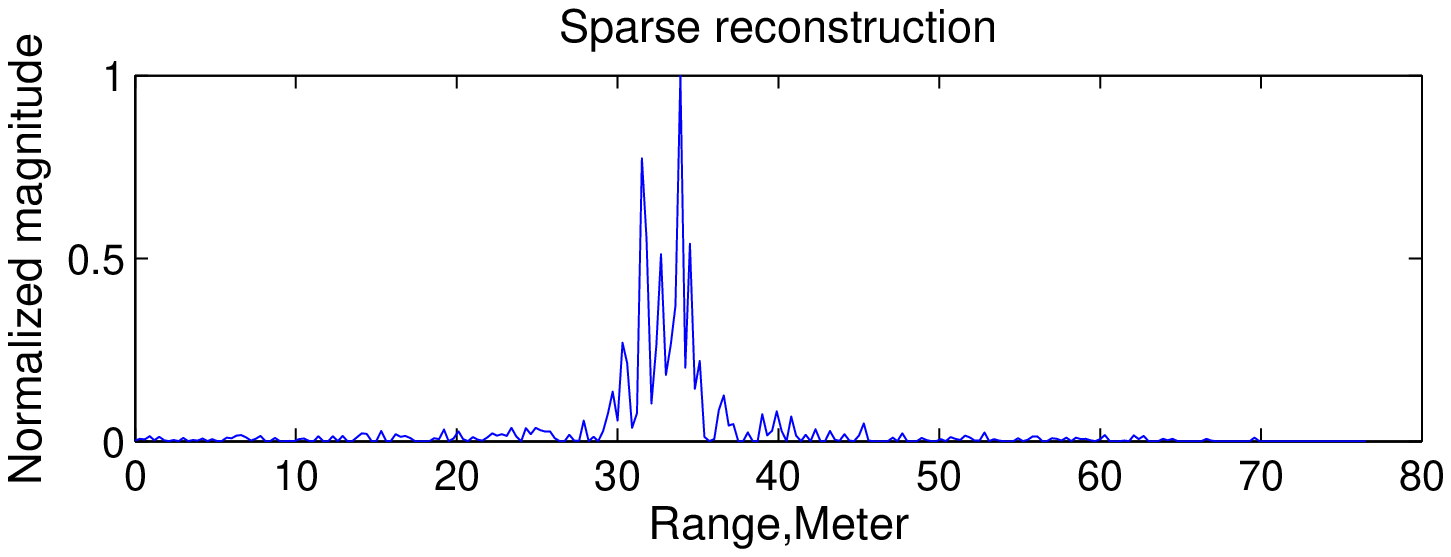}
} \caption[ The system model.] {Comparison of result obtained via
different methods for simulative data. (a) Model of scatters. (b) LS
result of missing data. (c) Sparse recovery of missing data.}
\end{figure}}

The results of simulation data obtained via different methods are
compared in Fig.1. Fig.1(b) show the result obtained from LS method
\cite{liu}, Fig.1(c) demonstrate the result from new approach. From
which it can be noted that LS method has created high sidelobe, and
the result by new method is more similar to original target range
profile. {\setlength{\abovecaptionskip}{0pt}
\begin{figure}[Similarity]
\centering%
\includegraphics[width=0.4 \textwidth] {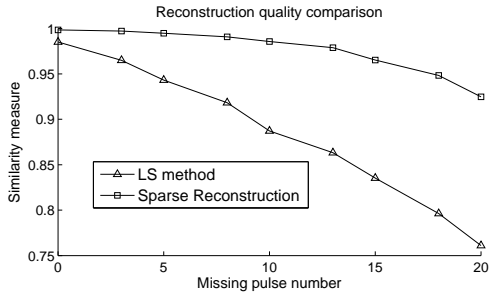}
\caption[ Similarity measure] {Similarity measure.}
\end{figure}}

To analyze the profiling results of the two methods quantitatively,
we measure the similarity between the simulated target and the
reconstruction profile by normalized cross correlation. Similarity
equals $1$ means perfect reconstruction. Fig.2 illustrate the
comparison. We increase the number of missing pulses from 0 to 20.
The line marked by ``$\vartriangle$'' denotes the similarity by LS
method, and line marked by ``$\square$'' denotes the similarity by
sparse recovery. Sparse recovery has an obvious advantage over the
LS counterpart.

\subsection{Real Radar Data}
We use real radar data obtained from SF radar. An experiment was
carried out in a wide and flat field. A single metal reflector was
placed 1010m away from the radar antenna. Experimental data of I/Q
channels was collected from the baseband of the radar receiver. All
parameters in the experiment were equal to the simulated data. We
discard 12 pulses randomly to simulate the missing data condition.
New approach is applied to the missing data. Profiling results are
compared to IDFT based LS method.
{\setlength{\abovecaptionskip}{-2pt}
\begin{figure}[]
\centering%
\subfigure []{
\includegraphics[width=0.4 \textwidth] {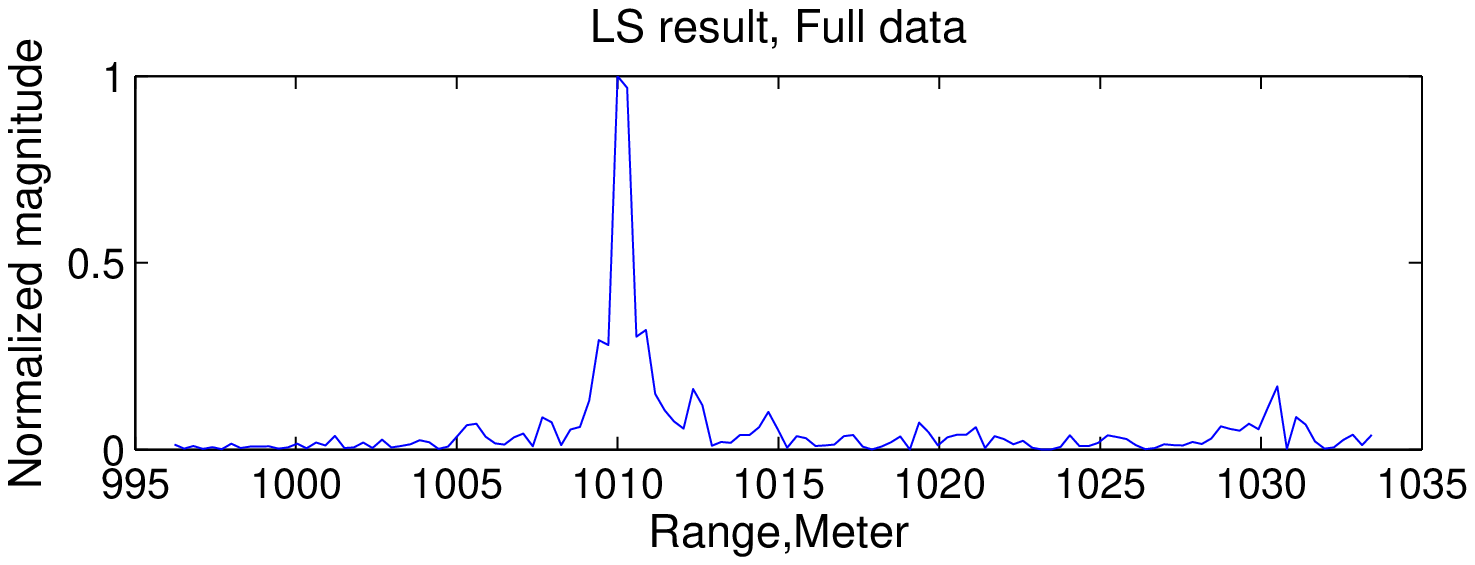}
} \subfigure []{
\includegraphics[width=0.4 \textwidth] {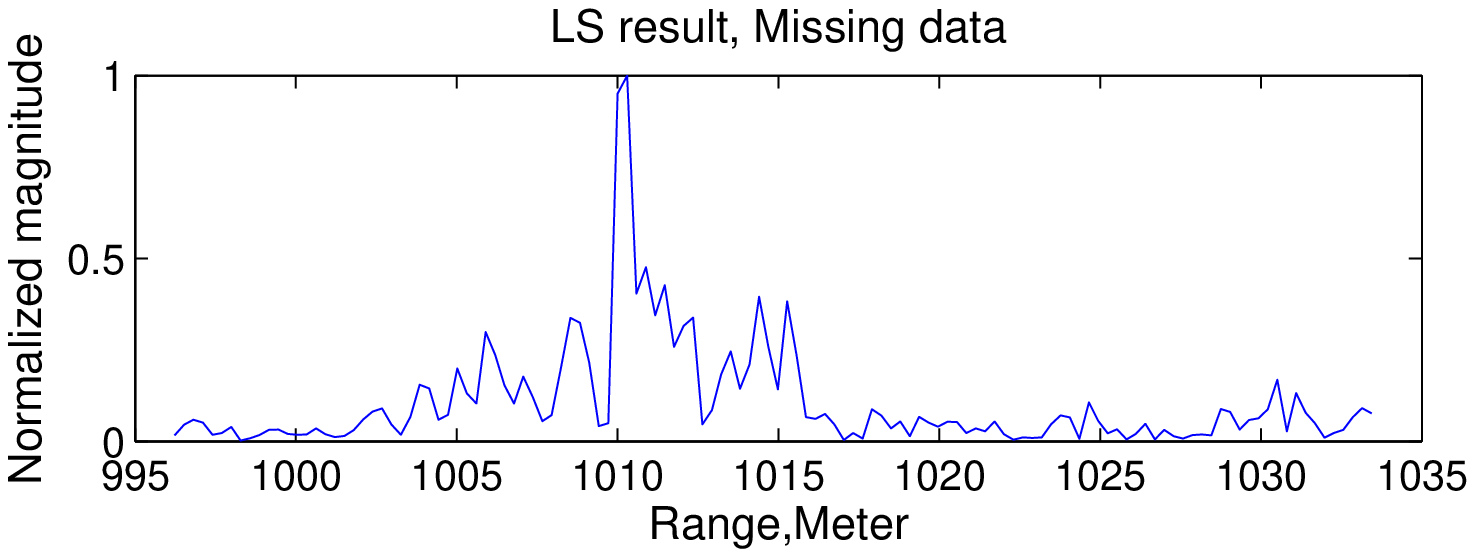}
} \subfigure []{
\includegraphics[width=0.4 \textwidth] {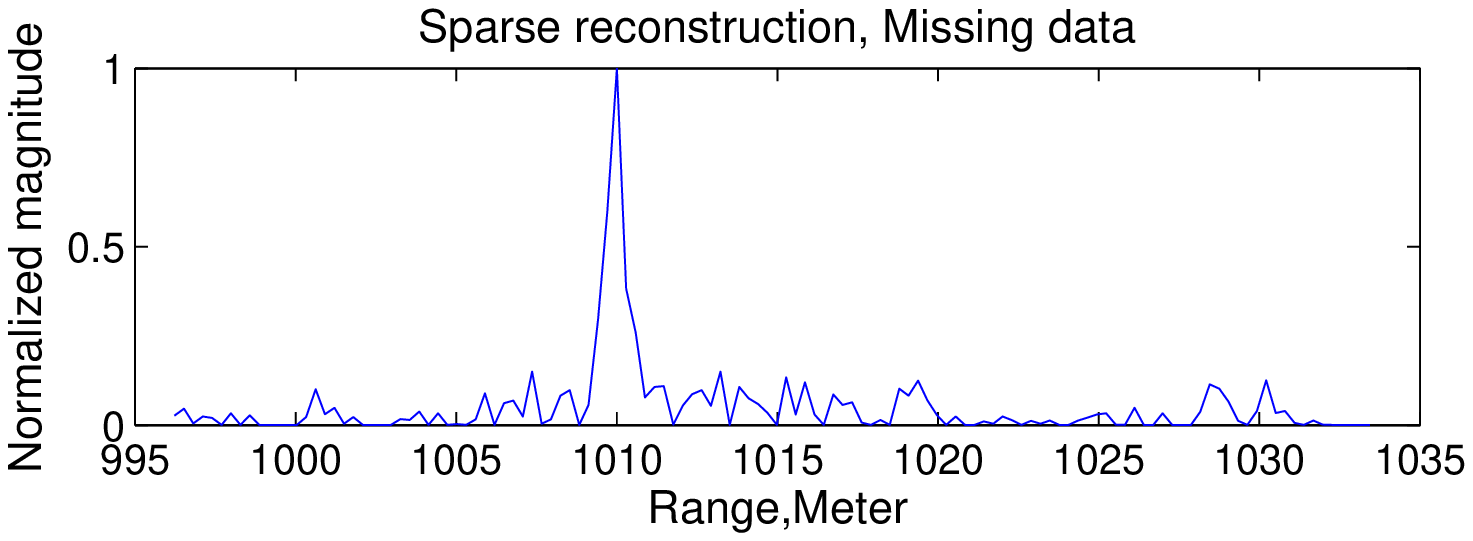}
} \caption[ The system model.] {Comparison of result obtained via
different methods for real radar data. (a) Full pulse recovery. (b)
LS result of missing data. (c) Sparse recovery of missing data.}
\end{figure}}

Fig. 3(a) shows the profiling result of LS method with full data.
Fig. 3(b) and 3(c) compare the profiling results to the missing data
by LS method and the new method respectively. LS method exhibits
high sidelobe as predicted, while the profiling result by new method
is similar to full data profiling. Sparse recovery outperforms LS
using real radar data.


\section{Conclusion}
The application of sparse recovery in extended HRR profiling for SF
radar is illustrated. The simulated data and real data experiments
prove that the proposed method is an appropriate tool to deal with
missing data problem. Profiling quality of the new method has an
obvious advantage over IDFT based least square method, if some
pulses are missing. Moreover, it can profile multiple
coarse-range-bins simultaneously, indicating a wide profiling range.
The profiling result is not corrupted by ghost images. Further work
should consider reducing computational load for real-time
implementations.





%

\end{document}